\documentclass{iau} 
\usepackage{graphicx}
\usepackage{natbib}
\bibpunct{(}{)}{;}{a}{}{,}

\title[The evolution of magnetic hot massive stars] 
{The evolution of magnetic hot massive stars: \\ Implementation of the quantitative influence of surface
magnetic fields in modern models of stellar evolution}

\author[Z. Keszthelyi, G. Wade \& V. Petit]   
{Zsolt Keszthelyi$^{1,2}$, 
 Gregg A. Wade$^1$
 \and V{\'e}ronique Petit$^{3,4}$}

\affiliation{$^1$ Department of Physics, Royal Military College of Canada, \\ PO Box 17000 Station Forces, Kingston, ON, K7K 0C6, Canada  \\ email: {\tt zsolt.keszthelyi@rmc.ca} \\[\affilskip]
$^2$ Department of Physics, Engineering Physics and Astronomy, Queen's University, \\ 99 University Avenue, Kingston, ON, K7L 3N6, Canada \\
$^3$ Department of Physics and Space Sciences, Florida Institute of Technology, \\ 150 W. University Blvd, Melbourne, FL, 32904, USA \\
$^4$ Department of Physics and Astronomy, University of Delaware, \\ Newark, DE, 19711, USA}

\pubyear{2017}
\volume{329}  
\jname{The Lives and Death-throes of Massive Stars}
\editors{J.J. Eldridge, eds.}

\begin{document}

%
%
%


\def\jnl@style{\it}
\def\aaref@jnl#1{{\jnl@style#1}}

\def\aaref@jnl#1{{\jnl@style#1}}

\def\aj{\aaref@jnl{AJ}}                   
\def\araa{\aaref@jnl{ARA\&A}}             
\def\apj{\aaref@jnl{ApJ}}                 
\def\apjl{\aaref@jnl{ApJL}}                
\def\apjs{\aaref@jnl{ApJS}}               
\def\ao{\aaref@jnl{Appl.~Opt.}}           
\def\apss{\aaref@jnl{Ap\&SS}}             
\def\aap{\aaref@jnl{A\&A}}                
\def\aapr{\aaref@jnl{A\&A~Rev.}}          
\def\aaps{\aaref@jnl{A\&AS}}              
\def\azh{\aaref@jnl{AZh}}                 
\def\baas{\aaref@jnl{BAAS}}               
\def\jrasc{\aaref@jnl{JRASC}}             
\def\memras{\aaref@jnl{MmRAS}}            
\def\mnras{\aaref@jnl{MNRAS}}             
\def\pra{\aaref@jnl{Phys.~Rev.~A}}        
\def\prb{\aaref@jnl{Phys.~Rev.~B}}        
\def\prc{\aaref@jnl{Phys.~Rev.~C}}        
\def\prd{\aaref@jnl{Phys.~Rev.~D}}        
\def\pre{\aaref@jnl{Phys.~Rev.~E}}        
\def\prl{\aaref@jnl{Phys.~Rev.~Lett.}}    
\def\pasp{\aaref@jnl{PASP}}               
\def\pasj{\aaref@jnl{PASJ}}               
\def\qjras{\aaref@jnl{QJRAS}}             
\def\skytel{\aaref@jnl{S\&T}}             
\def\solphys{\aaref@jnl{Sol.~Phys.}}      
\def\sovast{\aaref@jnl{Soviet~Ast.}}      
\def\ssr{\aaref@jnl{Space~Sci.~Rev.}}     
\def\zap{\aaref@jnl{ZAp}}                 
\def\nat{\aaref@jnl{Nature}}              
\def\iaucirc{\aaref@jnl{IAU~Circ.}}       
\def\aplett{\aaref@jnl{Astrophys.~Lett.}} 
\def\apspr{\aaref@jnl{Astrophys.~Space~Phys.~Res.}}
\def\bain{\aaref@jnl{Bull.~Astron.~Inst.~Netherlands}} 
\def\fcp{\aaref@jnl{Fund.~Cosmic~Phys.}}  
\def\gca{\aaref@jnl{Geochim.~Cosmochim.~Acta}}   
\def\grl{\aaref@jnl{Geophys.~Res.~Lett.}} 
\def\jcp{\aaref@jnl{J.~Chem.~Phys.}}      
\def\jgr{\aaref@jnl{J.~Geophys.~Res.}}    
\def\jqsrt{\aaref@jnl{J.~Quant.~Spec.~Radiat.~Transf.}}
\def\memsai{\aaref@jnl{Mem.~Soc.~Astron.~Italiana}}
\def\nphysa{\aaref@jnl{Nucl.~Phys.~A}}   
\def\physrep{\aaref@jnl{Phys.~Rep.}}   
\def\physscr{\aaref@jnl{Phys.~Scr}}   
\def\planss{\aaref@jnl{Planet.~Space~Sci.}}   
\def\procspie{\aaref@jnl{Proc.~SPIE}}   

\let\astap=\aap
\let\apjlett=\apjl
\let\apjsupp=\apjs
\let\applopt=\ao

\maketitle

\begin{abstract}

Large-scale dipolar surface magnetic fields have been detected in a fraction of OB stars, however only few stellar evolution models of massive stars have considered the impact of these fossil fields. We are performing 1D hydrodynamical model calculations taking into account evolutionary consequences of the magnetospheric-wind interactions in a simplified parametric way. Two effects are considered: i) the global mass-loss rates are reduced due to mass-loss quenching, and ii) the surface angular momentum loss is enhanced due to magnetic braking. As a result of the magnetic mass-loss quenching, the mass of magnetic massive stars remains close to their initial masses. Thus magnetic massive stars - even at Galactic metallicity - have the potential to be progenitors of `heavy' stellar mass black holes. Similarly, at Galactic metallicity, the formation of pair instability supernovae is plausible with a magnetic progenitor.


\keywords{stars: magnetic fields, stars: mass loss, stars: evolution}
\end{abstract}

\firstsection 
\section{Introduction}
Surface magnetic fields are detected in about 10\% of hot stars \citep{wade2016}, and are understood to be of fossil origin, likely remaining from the star formation history or the pre-main sequence evolution of the star \citep{donati2009,braithwaite2015}. These surface fields are known to form a magnetosphere due to the interaction with line-driven winds of hot stars \citep{babel1997}. This interaction has been extensively studied in the literature by means of magnetohydrodynamic simulations \citep{ud2002,ud2008,ud2009} and analytical studies \citep{owocki2004,bard2016,owocki2016}.  

Two major effects occur that influence both the mass loss and angular momentum loss from the star on a dynamical time scale. i) The magnetosphere channels the wind material along magnetic field lines leading to an infall of mass \citep{ud2002}. The confined plasma is thus deposited back to the stellar surface, and hence the effective mass-loss rates are reduced. ii) The magnetosphere reduces the surface angular momentum budget via Maxwell stresses, which leads to an efficient slow down of the surface rotation (magnetic braking). We incorporate and test these `surface' effects caused by fossil magnetic fields in the one dimensional hydrodynamical stellar evolution code Module for Experiments in Stellar Astrophysics, MESA \citep{paxton2011,paxton2013,paxton2015}. 

\section{Methods}
\subsection{Mass-loss quenching}
When surface magnetic fields are coupled to the line-driven stellar winds of hot massive stars, then these magnetic fields are capable of channeling the mass along the field lines (`flux tubes'). To quantify this interaction, \cite{ud2002} introduced the magnetic confinement parameter, $\eta_\star$ (cf. their Equation 20), which describes the magnetic field energy compared to the stellar wind kinetic energy at the stellar surface, 
\begin{equation}\label{eq:eta}
\eta_\star = \frac{B_{\rm eq}^{2}R_{\star}^{2}}{\dot{M}_{B = 0}v_{\infty}},
\end{equation}
where $B_{\rm eq}$ is the surface equatorial magnetic field strength, $R_\star$ is the stellar radius, $\dot{M}_{B = 0}$ is the mass-loss rate the star would have in absence of the magnetic field, and $v_{\infty}$ is the wind terminal velocity. For convenience, we use the polar surface field strength $B_{\rm p}$, considering that $B_{\rm p} = 2 \, B_{\rm eq}$ for a dipolar field. The mass-loss quenching effect due to surface magnetic fields occurs when the stellar wind, confined to flow along closed magnetic loops, ultimately shocks, stalls, and returns to the stellar surface. This effectively reduces the mass-loss rates. 
%
%
%
According to \cite{ud2002}, the equatorial radius of the farthest closed magnetic loop, that is the closure radius, $R_c$, in a magnetized wind with a dipolar geometry at the stellar surface is of the order of the Alfv\'en radius $R_A$,
\begin{equation}
\label{eq:rc}
R_c \sim R_\star + 0.7(R_A - R_\star) \, .
\end{equation}
The location of the Alfv\'en radius corresponds to the point in the magnetic equatorial plane where the magnetic field energy density equals the wind kinetic energy, that is $\frac{R_A}{R_\star} ~ \approx  0.3 + \left(\eta_\star +0.25\right)^{1/4}$.
%
%
With the obtained polar field strength $B_{\rm p}(t)$, the non-magnetic mass-loss rate, terminal velocity, and the stellar radius, we calculate the Alfv\'en radius. It is straightforward then to obtain a parameter describing the escaping wind fraction $f_B$.
 \begin{equation}
    \label{eq:f}
    f_B = \frac{\dot{M}_{\rm eff}}{\dot{M}_{B=0}} = 1 - \sqrt{1-\frac{R_\star}{R_c}}.
\end{equation} 
where $\dot{M}_{\rm eff}$ is the effective mass loss allowed by a magnetosphere. This parametric description accounts for the fraction of the stellar surface covered by open magnetic field loops. Along these loops, wind material can escape effectively, while along closed loops, material will fall back to the stellar surface.

We imposed magnetic flux conservation on the evolving models, 
\begin{equation}
 F \sim 4\pi R_\star^2 B_p = \mbox{constant}. \label{eq:flux}
 \end{equation}
As a consequence, as the star evolves and the stellar radius $R_\star$ increases, the surface magnetic field strength changes according to:
 \begin{equation}
    B_{p}(t) = B_{p,0} \left(\frac{R_{\star,0}}{R_\star(t)}\right)^2,
  \end{equation}
where $B_{p,0}$ and $R_{\star,0}$ correspond to the polar field and stellar radius defined at the zero age main sequence. 
The final adopted mass-loss rate is obtained by scaling the current time step mass-loss rates with the escaping wind fraction $f_B$ allowing for mass to escape only via open loops, such that:
\begin{equation}
\dot{M}_{\rm eff} = f_B  ~ \dot{M}_{B = 0}, \label{eq:fb}
\end{equation}
where $\dot{M}_{B=0}$ refers to the non-magnetic mass-loss rates according to any applicable scheme. In our main sequence models above 12.5 kK, we adopt the Vink rates \citep{vink2000,vink2001}. However, uncertainties related to this scheme (or any other scheme) will have an impact on the quantitative results since the magnetic mass-loss quenching is coupled to the wind scheme.  

\subsection{Magnetic braking}
We implemented magnetic braking in MESA; this will be discussed in a forthcoming publication (Keszthelyi et al., in prep). We followed the prescription derived by \cite{ud2009}, and thus adopted an additional source of angular momentum loss,
\begin{equation}
\frac{\mathrm{d} J}{\mathrm{d} t} = \frac{2}{3} \dot{M}_{B=0} \, \Omega \, R_{\star}^{2} \, [ (\eta_\star + 0.25)^{1/4} + 0.29]^{2},
\end{equation}
where $J$ is the total angular momentum, $t$ is the time, and $\Omega$ is the surface angular velocity. It is important to note that $\dot{M}_{B=0}$ in this equation strictly means that most of the angular momentum is lost by Maxwell stresses, and not by the effective mass loss. An important issue is that the way this surface angular momentum loss propagates into the interior layers is unknown, as the effect of large-scale fossil magnetic fields on differential rotation is not well constrained. \cite{meynet2011} have already implemented this prescription in the Geneva stellar evolution code, and tested two cases - solid body rotation and differential rotation. Indeed, the major problem arises from the fact that the surface angular momentum loss depends on the internal angular momentum transport mechanisms. While state-of-the-art stellar evolution modelling allows for testing both solid body and differential rotation, observational evidence on the internal rotational properties and thus on the transport mechanisms in hot stars only exists in very few cases (e.g. KIC 10526294 by \citealt{moravveji2015}, and V2052 Oph by \citealt{briquet2012}).

\section{Results}
\subsection{Models including mass-loss quenching} 
\begin{figure}[h]
\begin{center}
\includegraphics[width=3.6in]{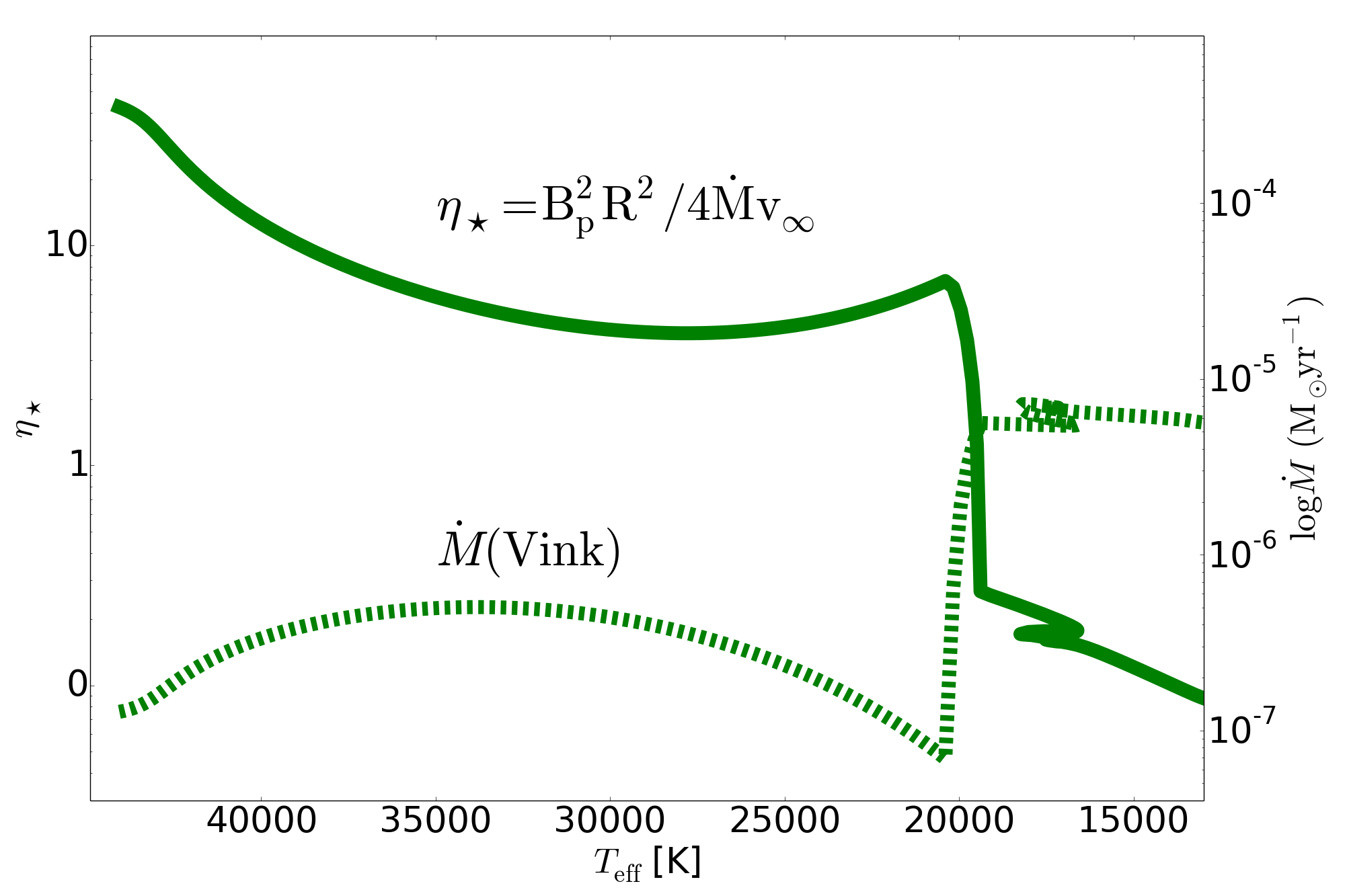}
 \caption{Shown is the evolution of the magnetic confinement parameter (solid line, left vertical axis, defined in Equation \ref{eq:eta}) and the effective mass-loss rates (dashed line, right vertical axis) against the effective temperature for a 40 $M_{\odot}$ Galactic metallicity non-rotating model.}
   \label{fig:fig3}
\end{center}
\end{figure}

We find that the evolution of the magnetic confinement parameter depends mostly on how $\dot{M}_{B=0} (T_{\rm eff})$ evolves, and in a general scenario the other parameters ($B_{\rm p}(t),R_{\star},v_{\infty}$, respectively) compensate each other. The impact of mass-loss rates on massive star evolution, and, in particular, the dependence of $\dot{M}_{B=0}$ on $T_{\rm eff}$ has recently been discussed in a more general context by \cite{keszthelyi2017}. In Figure \ref{fig:fig3} the evolution of magnetic confinement parameter (solid line, left ordinate) is contrasted to the evolution of the mass-loss rate (dashed line, right ordinate). Indeed, $\eta_\star$ is most sensitive to changes in $\dot{M}_{B=0}$. In particular, the large jump in mass-loss rate at the bi-stability (shifted down to 20 kK in accordance with the suggestion from \citealt{keszthelyi2017} based on findings from \citealt{petrov2016}) causes a drop in the magnetic confinement. This is powerful enough that the initially strong ($\eta_\star > 10$) and then moderate ($1< \eta_\star < 10$) magnetic confinement rapidly becomes weak ($\eta_\star < 1$).  
%

\subsection{The evolution of fossil magnetic fields}
\begin{figure}[h]
	\begin{center}
	\includegraphics[width=2.7in]{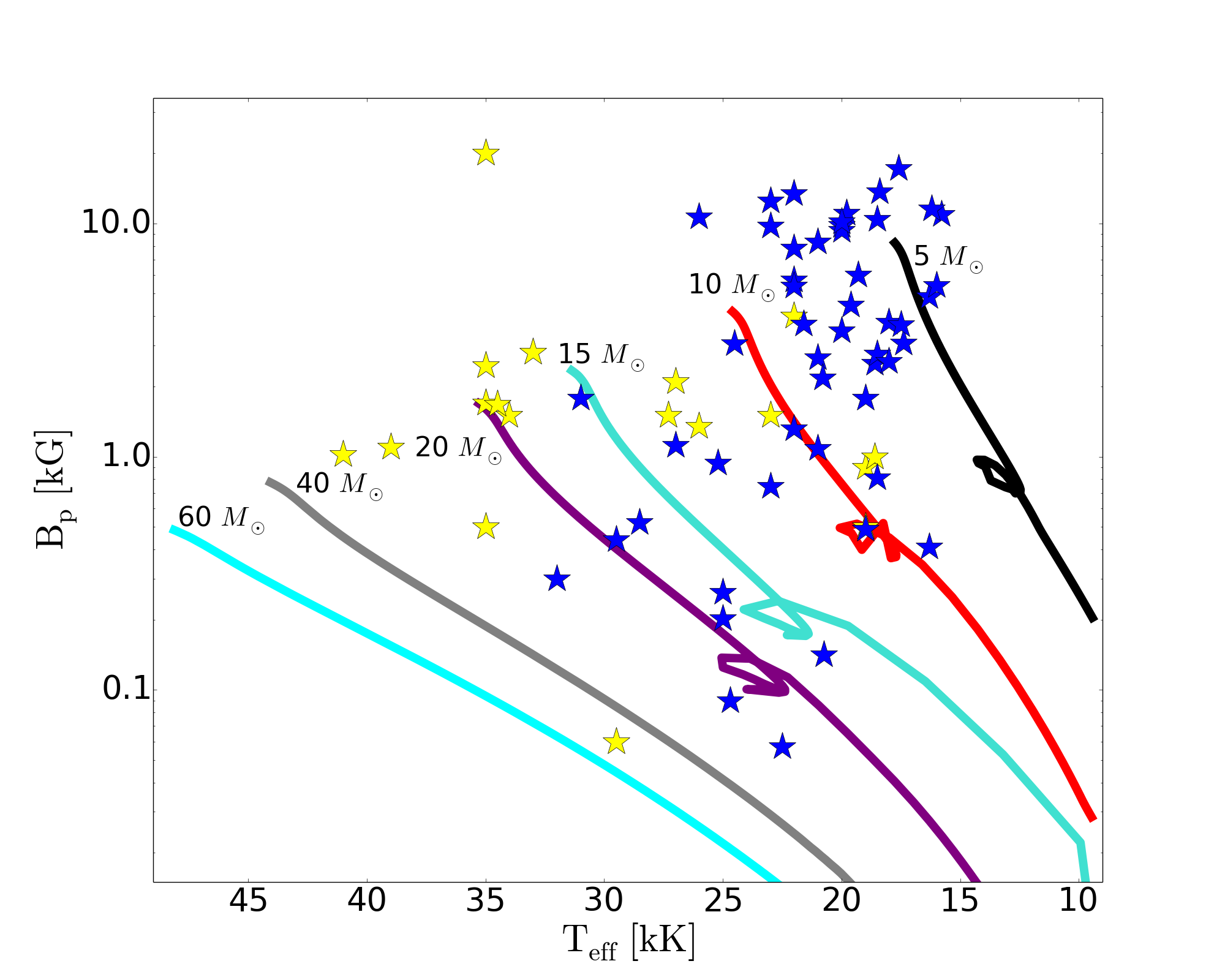}\includegraphics[width=2.7in]{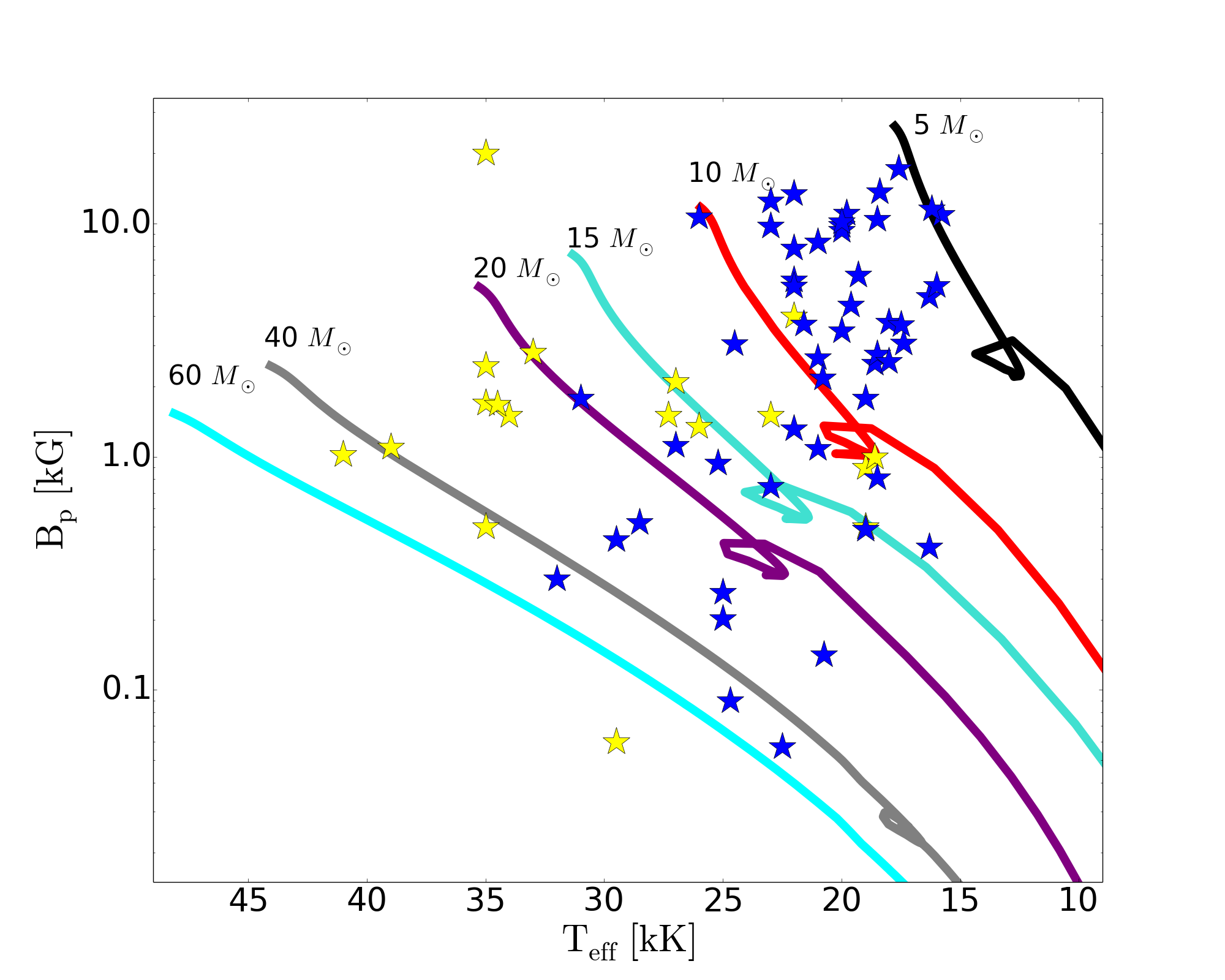}
	\caption{Plotted are the known sample of magnetic OB stars from \cite{petit2013} and Shultz et al. (in prep). The former are denoted with yellow stars, the latter with blue stars. Evolutionary models with different initial masses are plotted. \textit{Left}: Models adopting an initial magnetic flux of $F = 10^{27.5} \, \mathrm{G \, cm^2}$. This yields an initial dipole field strength in, e.g., the 40 $M_{\odot}$ model of $B_{\rm p} = 1 \, $ kG. \textit{Right}: Models adopting an initial magnetic flux of $F = 10^{28} \, \mathrm{G \, cm^2}$. This yields an initial dipole field strength in, e.g., the 40 $M_{\odot}$ model of $B_{\rm p} = 2.5 \, $ kG.}\label{fig:fieldevol}
	\end{center}
	\end{figure}

A key ingredient to incorporate the effects of fossil magnetic fields in stellar evolution models is currently poorly understood: the evolution of stellar magnetic fields over evolutionary timescales. Adopting magnetic flux conservation (Equation \ref{eq:flux}) allows for calculating the first grid of models that adopt a field evolution other than a constant magnetic field strength (that is, increasing flux) as in the work by \cite{meynet2011}. Two major scenarios are considered. The range of observed magnetic field strengths is very large, over 3 orders of magnitude in stellar remnants (e.g. \citealt{ferrario2006}). Interestingly, this range is consistent with the scale of field strengths in main sequence stars, and thus supports magnetic flux conservation. However, in light of theoretical considerations (e.g. \citealt{braithwaite2008, braithwaite2015}) a plausible scenario of magnetic flux decay should be taken into account. Most commonly, Ohmic decay is speculated to lead to magnetic flux decay. Additionally, observations may also argue for such a trend \citep{landstreet2007,fossati2016}.  
Figure \ref{fig:fieldevol} shows stellar evolution models adopting magnetic flux conservation to describe the field evolution. A sample of magnetic OB stars from \cite{petit2013} and Shultz et al. (in prep) are shown and their magnetic fluxes were calculated based on the measured magnetic field strengths and stellar radii. The MESA models are computed for two different initial flux values (left panel: $F = 10^{27.5} \, \mathrm{G \, cm^2}$, right panel:  $F = 10^{28} \, \mathrm{G \, cm^2}$). Since the initial flux is the same for all models, but the stellar radius is different due to the different initial masses, the initial field strength for higher mass (larger radius) models is smaller than that of lower mass (smaller radius) models.  

\section{Conclusions}

For the first time, we incorporated the effects of surface magnetic fields in the MESA code. These effects are based on the dynamical interactions between the magnetosphere and the stellar wind, and we conclude from our model calculations that they have a large impact on massive star evolution. Two evolutionary scenarios are discussed by \cite{petit2017} and Georgy et al. (in press).

\acknowledgements 

\scriptsize{We acknowledge the MESA developers for making their code publicly available. GAW acknowledges Discovery Grant support from the Natural Sciences and Engineering Research Council (NSERC) of Canada. VP acknowledges support provided by the NASA through Chandra Award Number GO3-14017A issued by the Chandra X-ray Observatory Center, which is operated by the Smithsonian Astrophysical Observatory for and on behalf of the NASA under contract NAS8-03060.}

\normalsize

\bibliography{thp}



\end{document}